# Topological phase transition and quantum spin Hall edge states of antimony few layers

Sung Hwan Kim[1,2], Kyung-Hwan Jin[2], Joonbum Park[2], Jun Sung Kim[2], Seung-Hoon Jhi[2] & Han Woong Yeom[1,2,*]

[1]*Center for Artificial Low Dimensional Electronic Systems, Institute for Basic Science, 77 Cheongam-Ro, Pohang 790-784, Korea*

[2]*Department of Physics, Pohang University of Science and Technology, 77 Cheongam-Ro, Pohang 790-784, Korea*

**While two-dimensional topological insulators (2D TI) initiated the field of topological materials, only very few materials were discovered to date and the direct access to their quantum spin Hall edge states has been challenging due to material issues. Here, we introduce a new 2D TI material, Sb few layer films. Electronic structures of ultrathin Sb islands grown on $Bi_2Te_2Se$ are investigated by scanning tunneling microscopy. The maps of local density of states clearly identify robust edge electronic states over the thickness of three bilayers in clear contrast to thinner islands. This indicates the topological edge states emerged through a 2D topological phase transition predicted between three and four bilayer films in recent theory. The non-trivial phase transition and edge states are**

---

* E-mail: yeom@postech.ac.kr

**confirmed for epitaxial films by extensive density-functional-theory calculations. This work provides an important material platform to exploit miscroscopic aspects of the quantum spin Hall phase and its quantum phase transition.**

Topological insulators (TI's) possess distinct properties, among which most important is the existence of spin helical Dirac fermions on their edges protected robustly. Such topological edge states (TES's) have been fully demonstrated through experimental studies[1–6] and grant attractive applications in transporting spin currents for spintronic devices and in providing key elements of topological quantum computers[7,8]. Two-dimensional (2D) TI's are particularly interesting in applications because of their merits in fabricating electronic or spintronic device structures with their intrinsically localized edge channels. Their TES's represent two spin polarized channels, which correspond to quantum spin Hall (QSH) edge states. The 2D TI phase was established in HgTe/CdTe and InAs/GaSb quantum well structures through transport measurements[9–12] and a ultrathin film of a 3D TI $Sb_2Te_3$ was suggested to fall into the 2D TI phase[13]. Edge channels of 2D TI's were accessed microscopically for quantum well structures with scanning SQIUD but with only micrometer resolution[14,15]. More recently, scanning tunneling microscopy/ spectroscopy (STM/STS) studies[16–22] claimed the truly microscopic and direct observation of nanometer scale TES's of 2D TI's for Bi single bilayer (BL) films in which the QSH phase was predicted earlier[23,24]. However, these Bi single layer films were realized on substrates with strong interactions, which leaves large ambiguity in the topological nature of their edge states [19,25,26].

In this work, we report the realization of the 2D QSH phase in another material based on epitaxially grown thin films and demonstrate its merit in the unambiguous microscopic observation of the TES. We note on the recent theoretical prediction of a series of topological quantum phase transitions in ultrathin Sb films, particularly the one from trivial to the semimetallic QSH phase between 3 and 4 BL thickness[27]. We investigate local electronic structures of ultrathin Sb films grown on $Bi_2Te_2Se$ by STM/STS and *ab initio* calculations. The initial growth of Sb on $Bi_2Te_2Se$ exhibits islands of various heights with well-ordered zigzag edges. Our spatially resolved $dI/dV$ (STS) measurements clearly identify strong and robust edge electronic states for 4 and 5 BL films in contrast to the 3 BL film. Our calculations clearly indicate the topological phase transition between 3 and 4 BL's and the topological non-trivial edge states for 4 BL's both without and with the interaction of the substate. A 2D TI based on a thin film with a well resolved TES is thus established.

## Results

**Sb film growth on $Bi_2Te_2Se$.** Figure 1a presents topographic STM image of $Bi_2Te_2Se$ where Sb islands of 1–5 BL heights are formed. At this growth condition, the single BL film or islands are rare and the portion of 2, 3, 4, and 5 BL areas are about 18, 18, 8, and 2 %, respectively. In particular, islands of 3, 4, and 5 BL heights have compact (Fig. 1b and c) and atomically well ordered structures as shown in Fig. 1d and e. The surface of 2 BL films is rather disordered with vacancies and clusters. These close-up images show unambiguously that the islands are (111) oriented with a lattice constant of about 4.0 Å, being consistent with the previous report[28]. Their well defined edges are perpendicular to the $[11\bar{2}]$ direction (parallel to  direction in the *k* space) and have the zigzag atomic structure.

**Local electronic structure of Sb(111) films on Bi$_2$Te$_2$Se.** In order to reveal electronic structures of ultrathin Sb films, we performed $dI/dV$ measurements on Sb islands, for example, for an island with 3, 4, and 5 BL-height films together (Fig. 2a). We take detailed $dI/dV$ curves along a line crossing two neighboring edges (the arrow within Fig. 2a). The result of this scan is shown in Fig. 2d. The noticeable features are strong spectral intensities at about −0.5 and +0.1∼0.25 eV in filled and empty states, respectively (horizontal dashed lines). These energies correspond well to the valence band edges of the substrate and the hybridized state between conduction bands of the substrate and the Sb film, respectively (Supplementary Fig. 1). The gradual downshift of these features evidences the charge transfer between the film and the substrate, which depends on the film thickness. The *n*-type doping of the Bi$_2$Te$_2$Se substrate was also observed for the Bi film growth[19]. Between these energies, 2D valence states of the Sb film appear as weak spectral features (more details found in Supplementary Fig. 1). Even within this line scan, one can notice the rather regular spatial modulation of these 2D electronic states of Sb films (and also for those at +0.1~0.25 eV) due to the quasi-particle interference induced by the presence of step edges (Supplementary Fig. 2). The interference pattern analysis, data not shown here, supports their origin in valence states of the Sb film.

In addition to the spectral features within flat parts of the films, one can easily notice that the electronic states are substantially modulated on the step edges (blue and green rectangles in Fig. 2d), in particular in empty states as indicated by arrows. As detailed in Fig. 2e and f, there exist unique electronic states on the edge sites at about +0.36 and +0.28 eV for 4 and 5 BL films, respectively. In the 2D lateral map of $dI/dV$ intensities at these energies, one can clearly see that those electronic states are strongly confined along the edges (Fig. 2b and c) within 2 nm (Supplementary Fig. 3). This edge state cannot be explained by the energy shift of the neighboring state at +0.1~0.25 eV within the interior of islands since those states do not merge into the edge state and no band bending toward the edge is noticed for all the other spectral features ( Supplementary Fig. 4).

We performed similar experiments for islands of 3 BL films on the same substrate. The cross-edge $dI/dV$ line scan measurement is shown in Fig. 3. The 2D plot of the $dI/dV$ measurement is depicted on Fig. 3b. Even though the $dI/dV$ measurement shows rich spectroscopic features, the edge spectrum does not change drastically. The $dI/dV$ spectra averaged for the interior and the edge of the 3 BL film confirm only marginal changes on the edges without a strong edge-localized feature (Fig. 3c). This indicates unambiguously that the edges of 3 BL and thicker films have distinct electronic properties (Supplementary Fig. 5).

**Topological nature of ultrathin Sb films.** According to the previous theoretical work for Sb films[27], the bands of ultrathin Sb film undergo the band inversion between 3 and 4 BL's, which indicates a topological phase transition. We detailed this result with our own calculation. We studied the band structure of free standing Sb films of various thickness as a function of the strength of the spin-orbit interaction. As shown in Fig. 4a, above the thickness of 4 BL, the band gap at the Brilluion zone center closes and reopens as the spin-orbit interaction is increased. This is directly related to the band inversion; the detailed band dispersion analysis shown in Fig. 4b and c indicates that the bonding and antibonding characters of the corresponding bands are switched at the full strength of the spin-orbit interaction. This topological phase transition changes the characteristics of edge states too. In Fig. 4f, we show the calculation of the electronic band dispersion of a 4 BL film. This film has the nanoribbon geometry for the top layer to feature step edges as shown in Fig. 4e. This geometry mimics well the island and edge structure of the experiment. At the step edge of the top layer of the 4 BL film, we can identify the edge state band. This band has a Dirac dispersion with its Dirac point at the $\bar{X}$ point, where the band gap is largest, and with its spin-split branches dispersing into valence and conduction bands separately at $\bar{\Gamma}$. This clearly evidences the nontrivial TES character (Fig. 4f and Supplementary Fig. 5). Comparing with the experimental STS spectra, we related the edge state peak of the experiment to the top of the edge state band (arrow in Fig. 4e) while there is some energy difference. We prove further that the topological nature of the 4 BL film does not change upon the interaction with the substrate (Fig. 4e and h) and the epitaxial case is equivalent to the film under an electric field (Fig. 4d and Supplementary Fig. 6). For a thinner film, the edge state of 2 BL's branches are spin-split but merge at $\bar{\Gamma}$ forming topologically trivial Rashba spin-split bands for both floating and epitaxial films (Supplementary Fig. 5).

**Discussion**

The QSH edge state is claimed to be directly accessible in nanometer scale for Bi single BL films[18,19]. However, on the substrate like $Bi_2Te_3$ and $Bi_2Te_2Se$, the interaction with the substrate is strong enough to close the band gap and the topological nature of the edge state is not clear enough. On the other hand, the step edge state of the Bi surface layer[29] was also claimed as the QSH edge state. However, the 2D TI phase of the Bi(111) surface layer cannot be justified and the observed edge state corresponds to the trivial Rashba spin-split state[26]. In the present case, the edge enhancement of the spectral feature is very much clear and the interfacial interaction is largely diminished for the top layer of 4 or 5 BL films (Supplementary Fig. 5). The robustness of the edge state over defects such as kinks is remarkable in the LDOS maps and spectra (Fig. 2 and Supplementary Fig. 2). Moreover, the absence of the edge state in thinner films corroborates the relationship of the edge state with the 2D topolglical phase transition. These two facts, the great sensitivity to the film thickness and the total insensitivity to defects, manifest the non-trivial nature of the edge states observed here. The present observation is parallel to a largely different and very recent approach to the QSH edge state, the STM study of the step edge of a weak 3D TI, which corresponds to a stack of 2D TI layers[29]. This and the present work would uniquely make it possible to investigate microscopic details of a 2D QSH TES. In addition, the present system would make it possible to investigate the 2D topological quantum phase evolution from trivial to nontrivial 2D and 3D phases, which occur as a function of the film thickness.

## Methods

**STM/STS measurements.** The STM/STS experiments were performed in ultrahigh vacuum better than $5 \times 10^{11}$ Torr, using a commercial low temperature STM (Unisoku, Japan) at ~78 K. STM topographic images were obtained using the constant current mode. The STS spectra (*dI/dV* curves) were acquired using the lock-in technique with a bias-voltage modulation of 1 kHz at 10-30 mV$_{rms}$ and a tunneling current of 500–800 pA.

**Sample growth.** Bi$_2$Te$_2$Se single crystals were used as the substrate, which were grown using the self-flux method[19,25]. The single crystals were cleaved *in vacuo* at room temperature. Ultrathin Sb islands were grown by a thermal effusion cell at room temperature.

**Theory.** *Ab initio* calculations were carried out in the plane-wave basis within the generalized gradient approximation for the exchange-correlation functional[30,31]. A cut-off energy of 400 eV was used for the plane-wave expansion and the *k*-points of 11×11×1 for the Brillouin zone sampling. In order to investigate edge states, we carried out calculations for a single BL nanoribbons of 15 Sb zigzag chains on top of flat and infinite 1 and 3 BL slabs. This reproduces step edges with the zigzag structure of 2 and 4 BL films. The Sb slabs and the nanoribbons were fully relaxed until the Helmann-Feynman forces were less than 0.01 eV/Å.

**Acknowledgements**

This work was supported by Institute for Basic Science (IBS) through the Center for Artificial Low Dimensional Electronic Systems (Grant No. IBS-R014-D1), and Center for Topological Matter (Grant No. 2011-0030046), and the Basic Science Research program (Grant No. 2012-013838) of KRF.

**Author contributions**

H.W.Y. conceived the project and organized the collaboration. S.H.K. performed the STM/STS measurements. K.-H.J. and S.-H.J. carried out first-principles calculations. The substrate was made by J.P. and J.S.K. S.H.K. and H.W.Y. wrote the manuscript with the comments from the other authors.

**Supplementary Information**

**Competing Interests**   The authors declare that they have no competing financial interests.

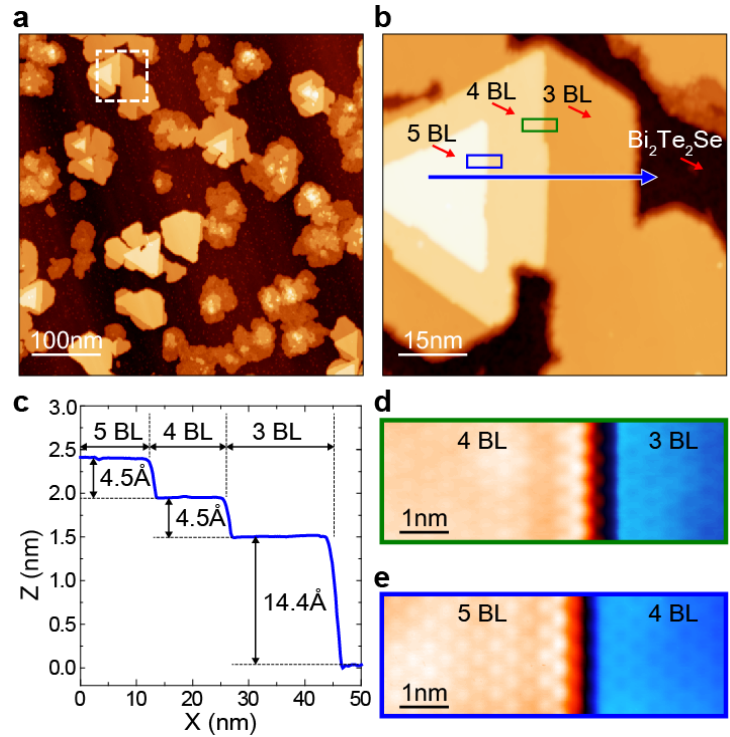

**Figure 1 Sb film growth on Bi$_2$Te$_2$Se.** (**a**) STM image for Sb films and islands on Bi$_2$Te$_2$Se (tunneling conditions of V$_s$= 1 V and I$_t$= 50 pA). Well ordered Sb islands appear over 2 BL's thickness. (**b**) Close-up STM image for an ordered Sb island, indicated by the dashed rectangle in (**a**). (**c**) The height profile of the island along the blue arrow in (**b**). The single layer height is about 4.5Å corresponding to one Sb(111) BL[28,32]. The step edge structures of (**d**) 4 and (**e**) 5 BL parts, the green and blue rectangles in (**b**) respectively. Zigzag chains of edge atoms are clearly revealed (V$_s$= 0.1 V, I$_t$= 1.5 nA).

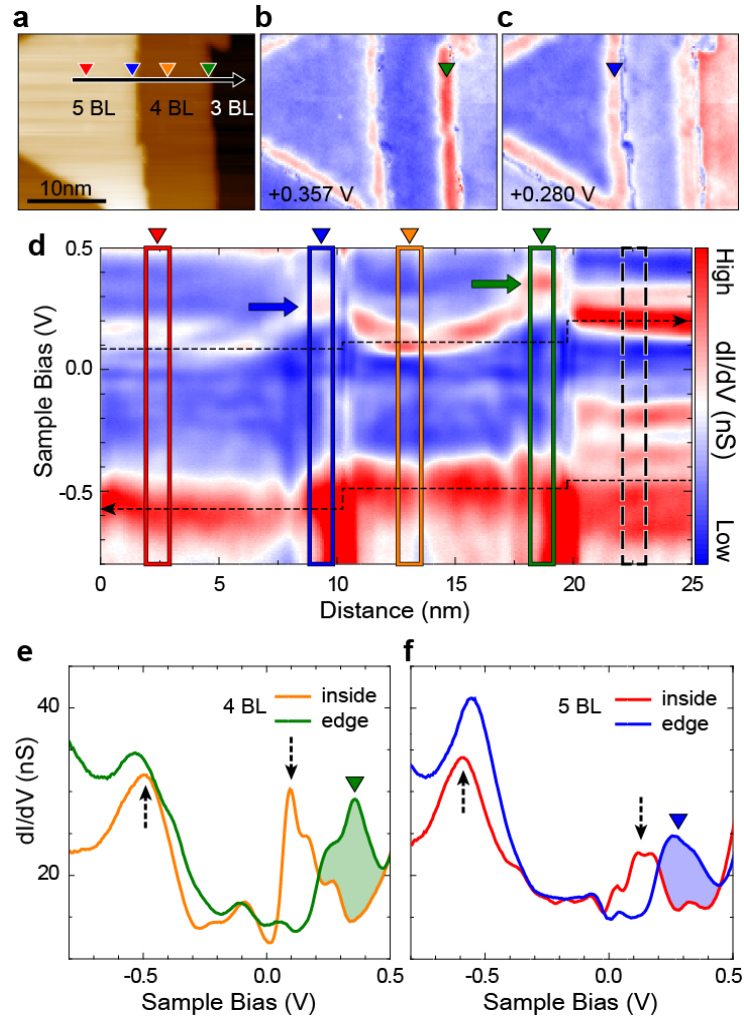

**Figure 2 Local electronic structures and edge states of Sb 4 and 5 BL.** (**a**) STM topography image (30 nm × 22.5 nm) on Sb island obtained simultaneously with the STS ($dI/dV$) measurements, which contains the edges of 4 and 5 BL films. The $dI/dV$ maps at two particular energies for edge states; (**b**) +0.357 V and (**c**) +0.280 V. (**d**) The 2D plot of the $dI/dV$ line scan measured along the black arrow indicated in (**a**), which crosses two zigzag edges. Averaged $dI/dV$ curves taken from the red, blue, orange, and green rectangles in (**d**), representing the center and edge parts of 4 and 5 BL films, are shown in (**e**) and (**f**), respectively. Around +0.36 and +0.28 eV, the strongly enhanced edge states appear [the green and blue arrows in (**d**) and arrow heads in (**e**) and (**f**)].

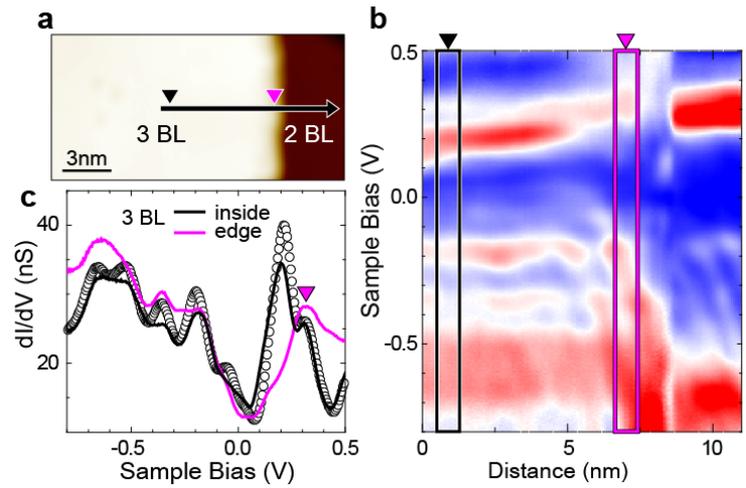

**Figure 3 Local electronic states of Sb 3 BL.** (**a**) STM topography image of a Sb island with its edge from a 3 BL film. (**b**) The 2D plot of the STS $dI/dV$ line scan along the black arrow in (**a**). (**c**) Averaged $dI/dV$ curves taken from the film and the edge (black and violet rectangles, respectively). The data in open circles are from a different 3 BL film shown in Fig. 2a (the dashed rectangle in Fig. 2d), exemplifying the consistency of the STS data over different islands and films.

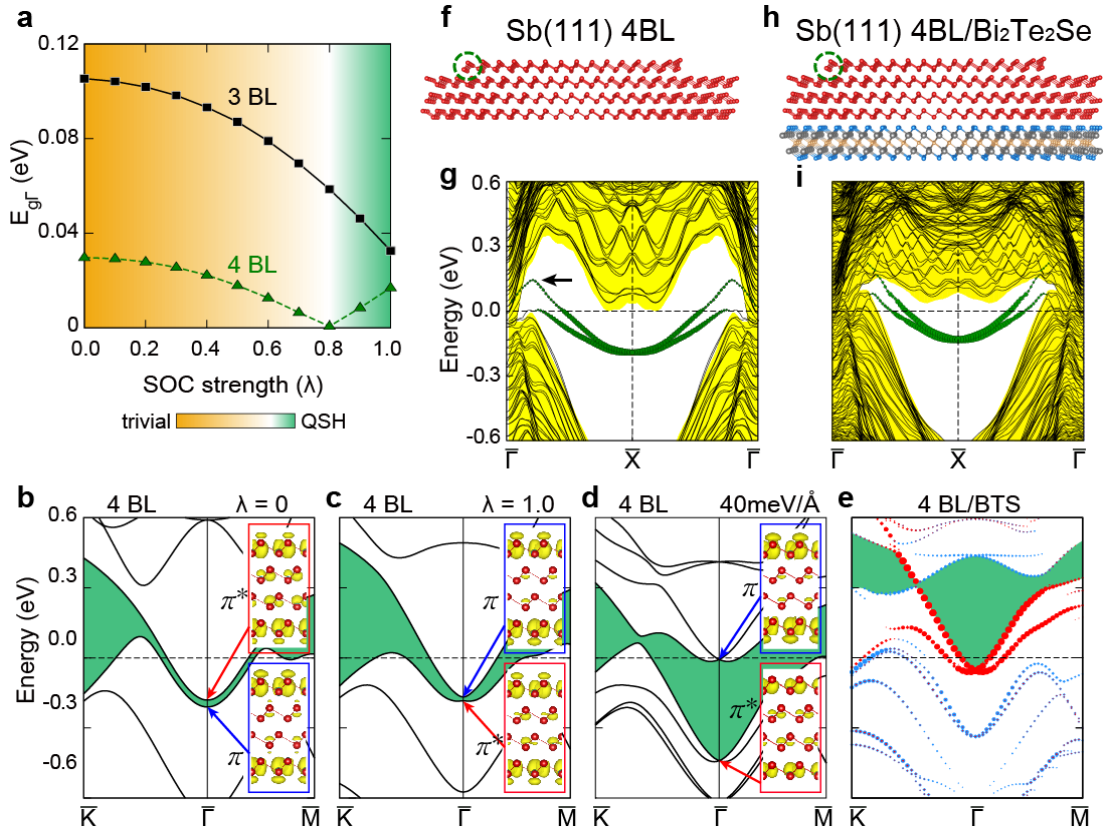

**Figure 4 *Ab initio* calculations for Sb films and nanoribbon.** (**a**) Calculated band gap at $\bar{\Gamma}$ ($E_{g\Gamma}$) with respect to the strength of spin-orbit coupling (SOC, $\lambda$) for 3 and 4 BL infinite Sb(111) films. The SOC strength ($\lambda$) is artificially set to partial fractions of the true value of SOC. The calculated band structures for Sb 4 BL are depicted for $\lambda$ of (**b**) 0 and (**c**) 1 along the $\bar{K}$-$\bar{\Gamma}$-$\bar{M}$ direction. The band inversion between $\pi$ and $\pi^*$ states ( insets ) occurs after the reopening of the band gap over $\lambda=0.8$. (**d**) The calculated band structure of Sb 4 BL when electric fields is applied artificially about 40 meV/Å. Even though the electric field exist, the band inversion is maintained. (**e**) Calculated band structure for Sb 4 BL films on $Bi_2Te_2Se$. The step edge structure of a 4 BL zigzag-edged Sb(111) nanoribbon (**f**) without [(**h**) with] the substrate. The top layer has a nanoribbon geometry to generate step edges. The calculated band structure of the 4 BL film with step edges along the $\bar{\Gamma}$-$\bar{X}$-$\bar{\Gamma}$ direction (**g**) without [(**i**) with ] the substrate. The band in green dots are localized on the edge atoms

[dashed circle in (**f**)]. The edge states bands have two branches dispersing out from the conduction and valance bands and crossing at $\bar{X}$ point to form a topologically non-trivial Dirac band.

*Supplementary Information for*

Topological phase transition and quantum spin Hall edge states of antimony few layers

Sung Hwan Kim,[1,2] Kyung-Hwan Jin,[2] Joonbum Park,[2] Jun Sung Kim,[2]
Seung-Hoon Jhi,[2] and Han Woong Yeom[1,2]

[1]*Center for Artificial Low Dimensional Electronic Systems, Institute for Basic Science,
77 Cheongam-Ro, Pohang 790-784, Korea*
[2]*Department of Physics, Pohang University of Science and Technology,
77 Cheongam-Ro, Pohang 790-784, Korea*

A comparison between the calculated band structure and the STS spectrum for a Sb 4 BL film on $Bi_2Te_2Se$. The rich features of the *dI/dV* curve represent well the band structure. The states indicated by blue dashed lines originate mainly from the band edges of the substrate, which are located near +0.1 and -0.5 V. The other smaller features around -0.10, +0.15, and +0.27 eV (the red dashed lines) correspond to the states of Sb films, which have some degree of hybridization with those of $Bi_2Te_2Se$.

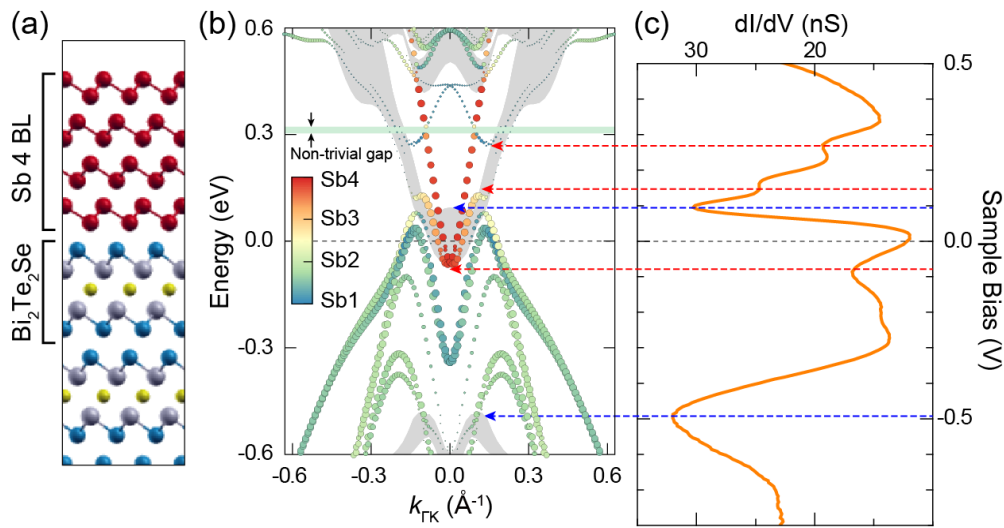

**Supplementary Figure 1. The origins of Sb 4 BL's STS spectral features.** (a) The atomic structure model for the *ab initio* calculation of Sb 4 BL's on $Bi_2Te_2Se$ (BTS). The structure is simulated by a supercell with Sb 4 BL's on one surface of a slab of six quintuple layers BTS and a vacuum layer of 20 Å between the cells. During structural relaxation, the atoms of Sb film and BTS surface 3 layers are allowed to relax until the forces are smaller than 0.01 eV/Å and the van-der Waals interaction is considered. (b) The calculated band structure for the Sb 4 BL film on BTS. Colors indicate which Sb layer the states originate from. The electronic states of BTS are within the grey region. (c) The STS spectrum measured inside of Sb 4 BL's.

Local density of states (LDOS) measurements for the 4 and 5 BL Sb films. The edge localized electronic states are very pronounced with well defined energies at +0.36 and +0.28 eV [(g) and (f)], for the 4 and 5 BL Sb films, respectively. In the range from +0.008 to +0.16 eV, the quasi-particle interferences within the films are clearly observed, which originate from the electron scattering of the surface states of the film by step edges [(b)-(e)]. The edge states are very robust against the defects so that their LDOS are affected only marginally on defects such as kinks [(h) and (i), the arrows in (f) and (g)].

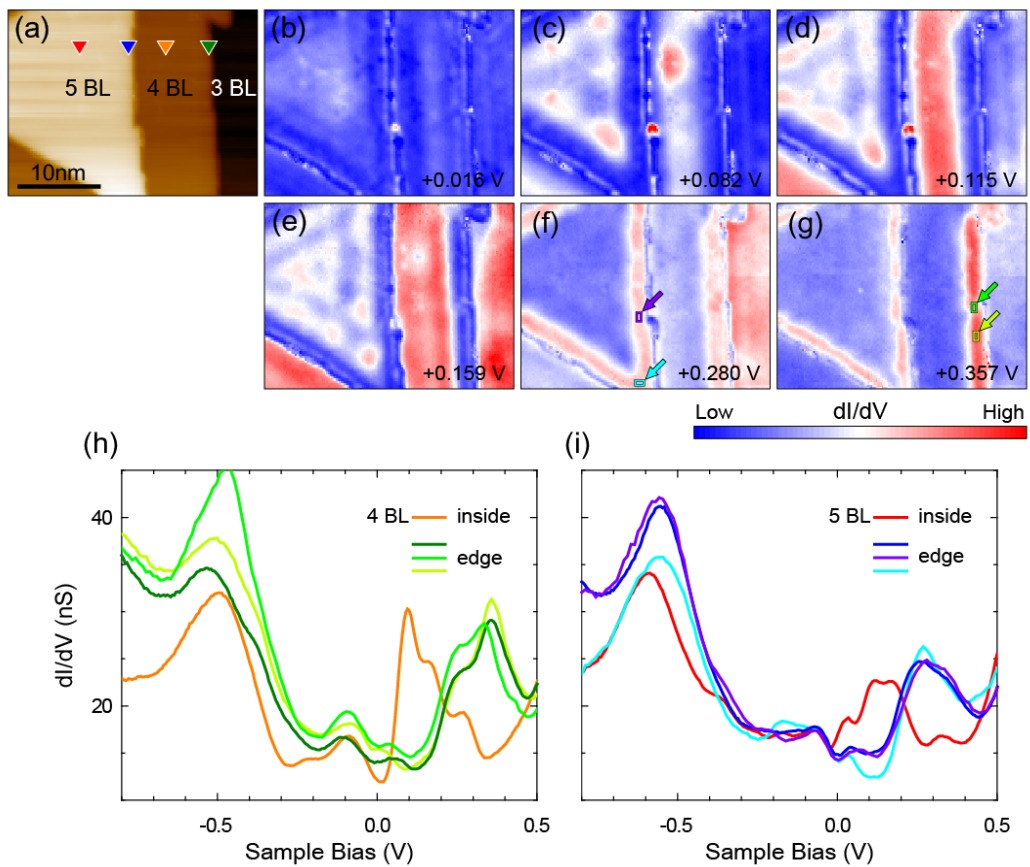

**Supplementary Figure 2. STS maps for several energies near Sb 4 and 5 BL edge.** (a) STM topographic image on a Sb island obtained simultaneously with the STS (*dI/dV*) maps. The *dI/dV* LDOS maps at several energies at (b) +0.016, (c) +0.082, (d) +0.115, (e) +0.0159, (f) +0.280, and (g) +0.357 eV. (h) and (i) *dI/dV* curves obtained along the edge channels but on defect sites indicated with colored boxes and arrows in (f) and (g) as compared with the *dI/dV* curves of Fig. 2e and f.

The STS (*dI/dV*) map of a Sb 4 BL film shows that the edge state distributes within about 2 nm from the edge. This matches well with the charge density plot of the edge states calculated for a Sb 4 BL nanoribbon.

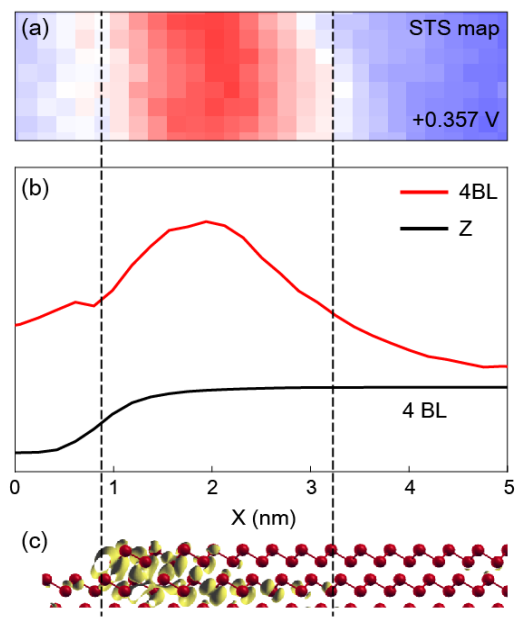

**Supplementary Figure 3. Spatial distribution of Sb 4 BL's edge states** (a) STS (*dI/dV*) map near the Sb 4 BL edge at +0.357 eV, which is taken from Fig. 2(b). (b) The averaged LDOS profile (red) of (a) together with the topographic line profile (black). (c) A charge density plot of the calculated edge states for a Sb 4 BL nanoribbon (the yellow blobs) , which corresponds to the state indicated by the arrow in Fig. 4(g).

The quasiparticle interference is observed near the Sb islands edges [Supplementary Fig. 2]. This explains the apparent energy shift of the state at +0.1 eV to about 0.25 eV toward the edge of 4 BL [middle of (a) and (c)]. Since the other states [black dashed line in (c)] have no such dispersion, this cannot be due to a band bending effect. The distinction between this dispersing state and the edge state [green triangles in (c)] is unambiguous.

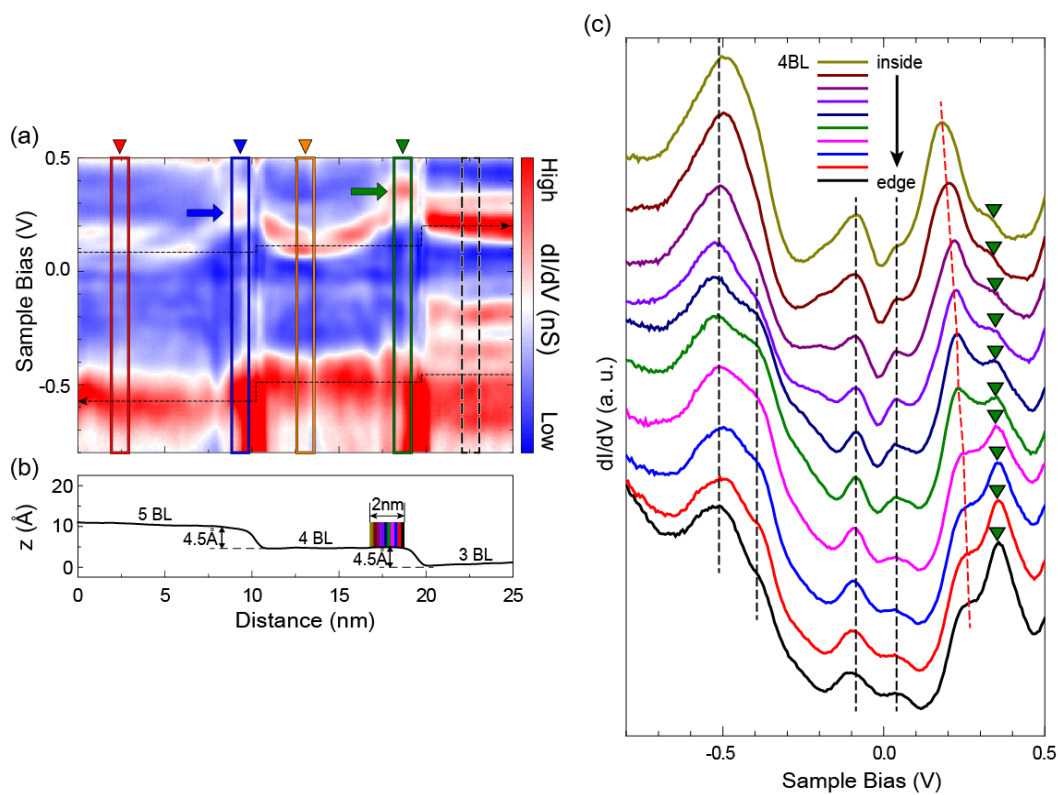

**Supplementary Figure 4. The STS measurements near the Sb 4 BL edge.** (a) The STS map crossing two step edges of 5 and 4 BL films, which is taken from Fig. 2d. (b) The topographic profile obtained during the STS measurements. (c) STS spectra near the Sb 4 BL edge. The line colors indicate the position where the STS obtained, which is indicated at (b). The energies of spectral features (the dashed lines) do not change except for one peak [the red dashed line], which is due the quasiparticle interference dispersion [Supplementary Fig. 2(b)-(e)].

*Ab initio* calculations for the edge states of Sb nanoribbon structures without and with the substrate. As shown in (a) and (b) the top layers have the step edge structures, which are fully relaxed. The trivial edge state with the Rashba spin splitting [2 BL case in (c)] and the non-trivial edge state with the Dirac dispersions [4 BL in (d)] are contrasted by their band dispersion near $\bar{\Gamma}$: In the case of 2 BL, the edge states converge into conduction bands. In contrast, the branches of the edge states of the 4 BL are separated into the conduction and the valance bands. If the spin-orbit coupling of the top layer of 4 BL is set to zero, the dispersion of the edge states become trivial merging into the conduction bands [(e)]. This difference is due to the topological phase transition and the band inversion occurring between 3 and 4 BL thickness. This nature is preserved even if the effect of the substrate is included [(f) and (g)].

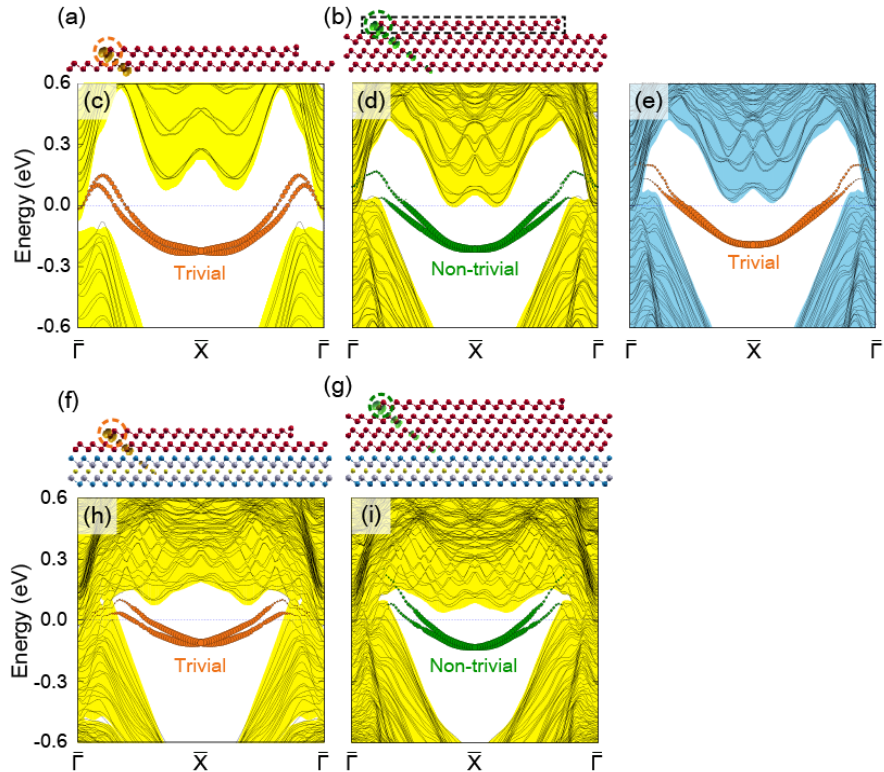

**Supplementary Figure 5. The topological nature of the edge states for each Sb nanoribbon.** Atomic structures used in the calculations for (a) Sb(111) 2 BL and (b) Sb(111) 4 BL in the floating geometry with zigzag step edges on the top layers. Calculated band structures of each model are depicted on (c) and (d) along the $\bar{\Gamma} - \bar{X} - \bar{\Gamma}$ direction. The bands represented by blue and green dots originate from step edges [the atoms within dashed circles in (a) and (b)], respectively, whose charge densities at $\bar{X}$ point are overlaid in the structure models. (e) The calculated band structure of the 4 BL nanoribbon with the spin-orbit coupling of the top layer [black dashed line in (b)] set to zero. (f) and (g) are structures for Sb(111) 2 BL and 4 BL films on $Bi_2Te_2Se$. The corresponding band structures are depicted in (h) and (i).

When the Sb 4 BL film is grown on $Bi_2Te_2Se$, the potential gradient is generated because of the presence of the substrate. Also the broken inversion symmetry causes bands to split. Nevertheless, *ab initio* calculations verify that the QSH phase is preserved. This can be seen by the evolution of the calculated band structures with the distance between the Sb film and $Bi_2Te_2Se$ varied to tune the substrate effect [(g)-(j)]. When the distance is set to the equilibrium position [(j)] with the full strength of the substrate effect, the band inversion and a non-trivial band gap are maintained. This substrate effect can be well understood by the electric field effect [(a)-(d)]. When the electric field is applied gradually, the bands split with the band inversion maintained. The band structure at 40 meV/Å is fully consistent with that with the substrate in (j) except for the rigid energy shift due to the doping. These results clearly prove that Sb 4 BL/$Bi_2Te_2Se$ is a QSH system.

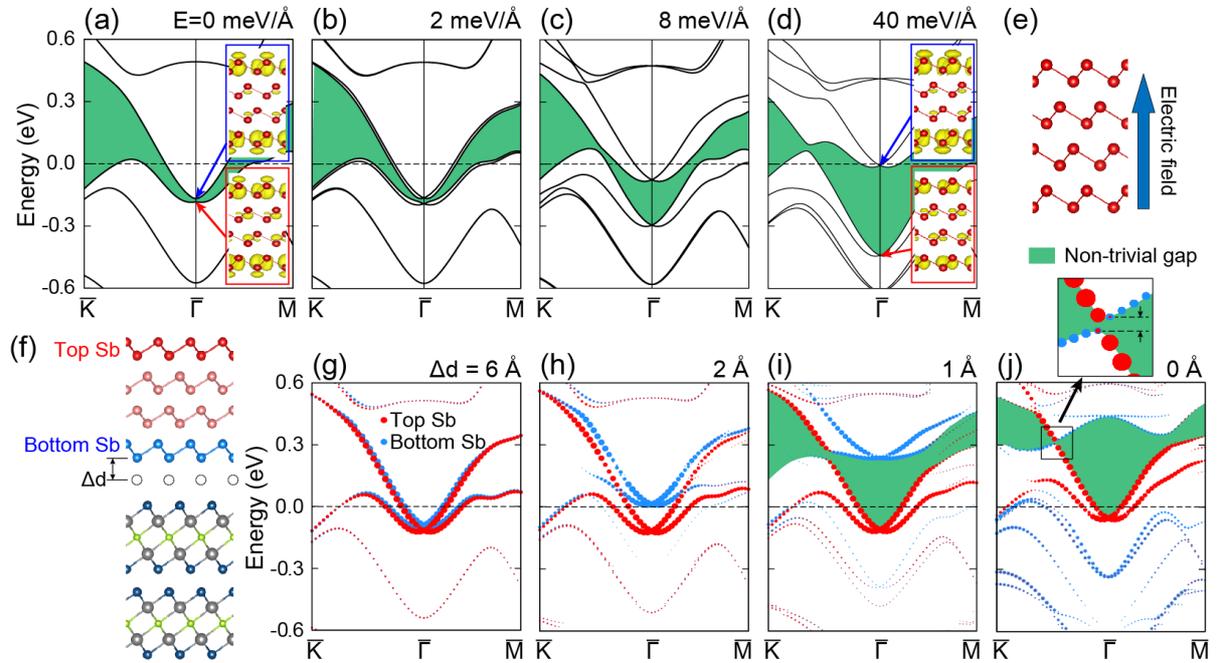

**Supplementary Figure 6.** *Ab* **initio calculation for Sb 4 BL when the potential gradient or the substrate exists.** (a)-(d) The calculated band structures when the electric field is applied with a strength of 0, 2, 8, and 40 meV/Å for a Sb 4 BL film as shown in (e). (f) Atomic structure of a Sb 4 BL on top of $Bi_2Te_2Se$. (g)-(j) calculated band structures with the substrate at a distance of 6, 2, 1, and 0 Å from the equilibrium position, respectively. The inset of (j) enlarges the non-trivial band gap of Sb 4 BL/$Bi_2Te_2Se$.